\documentclass{webofc}
\usepackage[varg]{txfonts}   
\usepackage{hyperref}

\begin{document}
\title{Fast Reconstruction and Data Scouting}
\author{\firstname{Javier}
  \lastname{Duarte}\inst{1}\fnsep\thanks{\email{jduarte1@fnal.gov}} \href{https://orcid.org/0000-0002-5076-7096}{orcid.org/0000-0002-5076-70960}}

\institute{Fermi National Accelerator Laboratory}

\abstract{Data scouting, introduced by CMS in 2011, is the use of specialized data streams based on reduced
  event content, enabling LHC experiments to record unprecedented
  numbers of proton-proton collision events that would otherwise be rejected by the
  usual filters. These streams were created to maintain sensitivity to
  new light resonances decaying to jets or muons, while
  requiring minimal online and offline resources, and taking advantage
  of the fast and accurate online reconstruction algorithms of the
  high-level trigger. The viability of this technique was demonstrated
  by CMS in 2012, when 18.8 fb$^{-1}$ of collision data at $\sqrt{s} =
  8$~TeV were collected and analyzed. For LHC Run 2, CMS, ATLAS, and LHCb implemented
  or expanded similar reduced-content data streams, promoting the
  concept to an essential and flexible discovery tool for the LHC.} 

\begin{flushright}
  FERMILAB-CONF-18-370-CMS-E-PPD
\end{flushright}

\maketitle

\section{Introduction}
\label{intro}
The Large Hadron Collider (LHC) provides proton-proton bunch crossings
with a center-of-mass energy $\sqrt{s} = 13~\mathrm{TeV}$ at a maximum
rate of 40 MHz. As such, the digital readout of the LHC detectors generate an enormous amount of raw data per bunch crossing; however, full
reconstruction of all collision events is not feasible with existing computing
resources necessitating the use of a trigger system. 

At the LHC, events of interest are selected using a two-tiered trigger
system~\cite{Khachatryan:2016bia,Ruiz-Martinez:2133909}. The first level (L1), composed of
custom hardware processors, uses information from the calorimeters and
muon detectors to select events at a rate of around 100~kHz
within a time interval of 2--4~$\mu\mathrm{s}$. The second level, known as
the high-level trigger (HLT), consists of a farm of processors running
a version of the full event reconstruction software optimized for fast
processing, and reduces the event rate to around 1~kHz before
data storage.

To make a trigger decision for each event, the HLT performs a
real-time (online) physics object reconstruction and applies a selection based on the
characteristics of the reconstructed objects. The full HLT algorithm
is a collection of independent trigger paths, each intended to serve a
different event selection purpose. Trigger paths consist of
sequences of producer modules, which build collections of objects; and
filter modules, which reject events that do not fulfill certain
criteria. Most trigger paths contain multiple phases of reconstruction
and filtering, and generally the later phases of reconstruction yield
physics objects whose performance is closer to that of their
offline counterparts. If an event is accepted by the final filter of
any trigger path, it is accepted by the HLT.

A number of factors restrict the trigger rates that the HLT can
achieve:
\begin{itemize}
\item The amount of data storage space and the maximum throughput of
  the data acquisition system (DAQ)
\item The capacity of the prompt reconstruction system
\item HLT computing resources, which limit the complexity of the online
reconstruction
\end{itemize} 

Data parking in CMS, or delayed stream in ATLAS, in which selected events are saved directly to tape with
no prompt reconstruction, has also been used in past LHC runs to increase
the amount of data collected. While this strategy circumvents the
restriction on rates imposed by the offline reconstruction system, it
is limited by the other factors listed above.

In this note, we describe data scouting, a technique that leverages
online reconstruction of physics objects in order to attain extremely
high trigger rates. Scouting complements data parking by providing new
opportunities for physics analysis outside the boundaries of the
traditional trigger strategy. 

Data scouting was first introduced by the CMS collaboration in
2011~\cite{CMS-DP-2012-022,CMS-PAS-EXO-11-094,Anderson:2016ron,Mukherjee:2017wcl}. The LHCb
and ATLAS collaborations implemented similar data-taking streams
for LHC Run 2~\cite{Aaij:2016rxn,ATLAS-CONF-2016-030,Aaboud:2018fzt}.

\section{Data scouting: an alternative trigger paradigm}

The aim of data scouting is to record physics events at the highest possible rate
while providing physics objects whose performance is suitable for
offline analysis. To do this, LHC experiments take advantage of the online
reconstruction algorithms. Trigger-level physics objects
are slightly less performant than their offline counterparts, but for certain
analyses, the difference does not significantly affect the
sensitivity. Saving only the trigger-level objects instead
of the full raw detector data makes the throughput and the size on
disk much smaller.

The data scouting strategy is implemented via dedicated streams at the
HLT. Each data scouting stream contains a number of trigger paths
(scouting triggers), which perform event reconstruction and selection in the same way as standard HLT paths. However, the
selection criteria are much looser than for standard paths and thus
the rate of events passing the selection is much greater.

For events passing one or more scouting triggers, additional online
reconstruction sequences are run in order to produce all physics
objects necessary for an offline measurement or search. The produced
objects are converted to a special compact event format and saved to
disk. The data recorded by scouting triggers is made available
offline and can be used for physics analysis. For these events, no
offline reconstruction is performed, and the raw data is not saved.

The scouting approach has the following advantages over the standard
trigger strategy:
\begin{itemize}
\item The reduced, compact event format requires negligible space on disk
and does not place any additional strain on the DAQ system. Events are
100 to 1000 times smaller on disk than the standard raw data format. 
\item No offline reconstruction is required; all reconstruction is performed online
\item Scouting trigger paths can run `in the shadow' of standard HLT
  paths, saving physics objects, reconstructed by the standard HLT
  paths even for events that are rejected by those paths
\end{itemize}
Scouting has been used to increase the total number of LHC events
recorded for physics by a factor of 2--6 beyond what the standard
trigger strategy provides.

\section{Data scouting in CMS}

The first scouting trigger was deployed at the CMS HLT during the last
few proton-proton fills of the 2011 LHC run. The trigger and associated stream
collected data equal to 0.13 fb$^{-1}$. Events with $H_\mathrm{T}$ (defined as the
scalar sum of jet transverse momenta) larger than 350~GeV were
recorded and saved in a reduced format, containing only the set of
anti-$k_t$ $R=0.5$ jets~\cite{Cacciari:2008gp,Cacciari:2011ma} reconstructed from particle-flow (PF) candidates by
the HLT. The data were used to perform a search for heavy resonances
decaying to dijets~\cite{CMS-PAS-EXO-11-094}. The search demonstrated
sensitivity to resonances with masses between 0.6 and 0.9 TeV, a
parameter region inaccessible to the standard CMS dijet resonance
search.

Subsequently, the strategy was repeated for the full 2012 CMS
dataset, lowering the scouting trigger selection to $H_\mathrm{T}$ >
250~GeV to accommodate an even larger rate of events. Due to CPU
concerns related to the high rate, calorimeter jets were reconstructed and
saved instead of PF jets. The collected data, corresponding to
18.8~fb$^{-1}$, were used to perform another dijet resonance
search~\cite{Khachatryan:2016ecr}, whose results were interpreted as
limits on the mass and coupling of a hypothetical leptophobic
$\mathrm{Z}^{\prime}$ resonance decaying to quarks. 

In Run 2, the data scouting strategy was expanded, with an aim to
maintain the ability to search very low in $H_\mathrm{T}$ using
calorimeter jets, while also providing an event format capable of
supporting a broader range of scouting analyses.
Two streams were deployed at the HLT for data taking in 2015--2018:
one saving an event content based on calorimeter jets (the
calo-scouting stream) and one saving an event content based on
particle-flow (PF) jets (the PF-scouting stream). The PF
algorithm~\cite{CMS-PRF-14-001} aims to reconstruct and identify each individual particle in an event, with an optimized
combination of information from the various elements of the CMS
detector. The PF jet energy resolution amounts typically to 15\% at
10~GeV, 8\% at 100~GeV, and 4\% at 1~TeV, to be compared to about 40\%, 12\%, and 5\% for
calorimeter jets. In order to run the PF algorithm online within
the allotted CPU time of around $200$~ms, a speed-optimized
configuration of the full CMS track reconstruction is used~\cite{TOSI20162494}.

Finally, new dedicated scouting event formats were developed, featuring a set of minimalist C++
objects to store scouting physics objects as vectors of basic data
types, ensuring low overhead and forward compatibility with future
versions of CMS software.

\subsection{Calo and PF-scouting streams}
\begin{figure}[htbp]
\centering 
\includegraphics[width=0.3\textwidth,clip=true,viewport = 400 200 624 600]{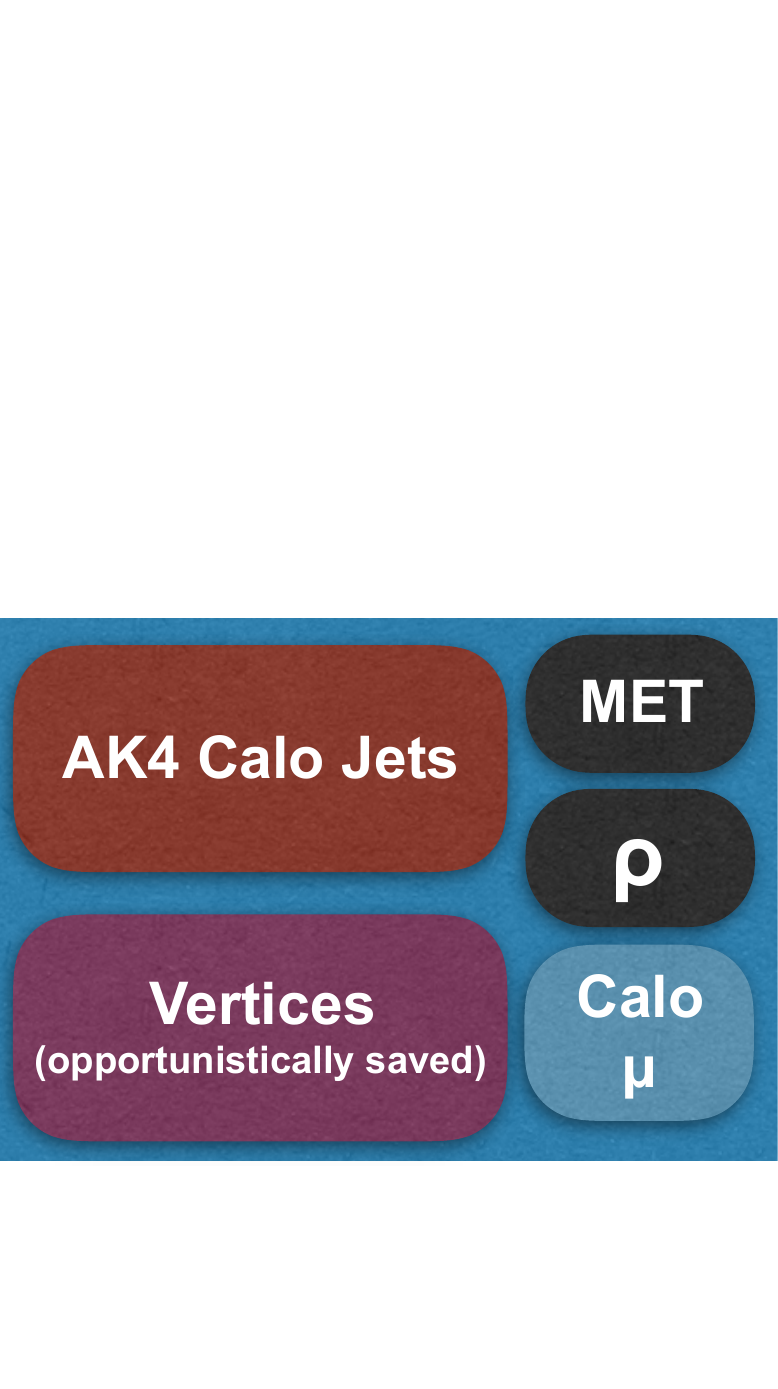}\hspace{0.1\textwidth}
\includegraphics[width=0.3\textwidth,clip=true,viewport = 400 200 624 600]{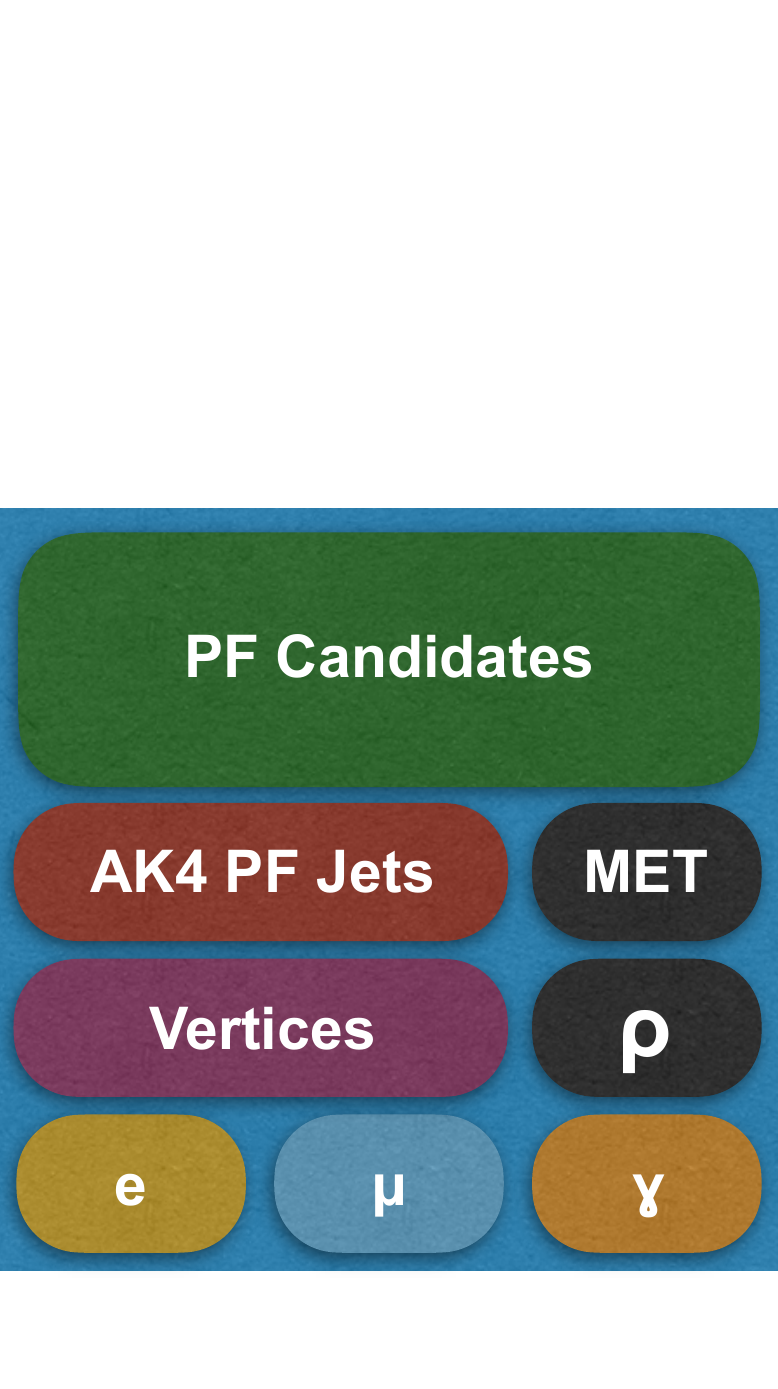}
\caption{Calo and PF-scouting event content. The corresponding sizes
  per event are approximately 1.5--3 kB and 10--15 kB, respectively}
\label{fig:content}
\end{figure}

Triggers in the calo-scouting stream reconstruct jets from calorimeter
deposits and the main signal trigger requires $H_{\mathrm{T}}>250$
GeV. The event content for this stream includes the reconstructed
calorimeter jets, the missing transverse momentum (MET), and $\rho$,
a measure of the average energy density in the event, as depicted in Fig.~\ref{fig:content}. Local pixel
track reconstruction provides b-tagging information for the jets.
The size of this event content is about 1.5--3~kB on average. 

Similarly, triggers in the PF-scouting stream run the online version of the full PF sequence to reconstruct
selected events and the main signal trigger selects events with $H_{\mathrm{T}}>410$ GeV. 
Additionally, the stream contains a trigger path selecting events with
two muons having invariant mass above 3 GeV, both with and without a
primary vertex constraint, allowing the possibility of searches for
prompt and displayed dimuon resonances. The event content for this
stream includes the reconstructed PF jets, the PF MET, the average energy density in the event $\rho$, a collection 
of primary vertices, and all PF candidates with $p_\mathrm{T}>0.6$
GeV. It also contains electron, muon, and photon objects, as depicted
in Fig.~\ref{fig:content}. The size of this event content 
is approximately 10--15~kB per event on average. 

Auxiliary prescaled trigger paths are also included both in calo and PF streams in order to facilitate
measurements of the signal trigger efficiency. For example, the
measured trigger efficiency of the calo-scouting stream as a function
of dijet invariant mass is shown in Fig.~\ref{fig:triggers}. 

\subsection{Monitoring stream}
To facilitate comparisons of the online physics objects to their
offline reconstructed counterparts, and to use the CMS Data Quality
Monitoring (DQM) framework to monitor the scouting data, a separate
monitoring stream was deployed. This stream contains prescaled versions
of all scouting triggers in the calo-scouting and PF-scouting
streams. Events selected for this stream are both saved in the reduced scouting
event format and sent to the CMS prompt reconstruction
system for offline processing, which enables detailed
object-by-object comparisons of the online and offline
performance. Representative rates and bandwidths for all CMS scouting and monitoring
streams in 2016 are shown in Tab.~\ref{tab:rates}.

\begin{table}[t]
\begin{center}
\begin{tabular}{lcc}  
\hline
Data stream    &    Rate at 10$^{34}$ cm$^{-2}$ sec$^{-1}$  &  Bandwidth (MB/s) \\ \hline
Calo-scouting $H_{\mathrm{T}}$ signal   &    3700                                                     &    11  \\
PF-scouting $H_{\mathrm{T}}$  signal     &    720                                                       &    9  \\
PF-scouting dimuon  signal    &   480                                                        &    6  \\
Commissioning   (PF + Calo)           &    30                                                         &    $<$1 \\
Monitoring                       &    26                                                         &   23 \\ \hline
\end{tabular}
\caption{CMS rates and bandwidths measured in 2016 data for various
  scouting data streams. The row marked `Commissioning (PF + Calo)'
  represents all auxiliary (non-signal) trigger paths in either the
  calo-scouting or PF-scouting streams.}
\label{tab:rates}
\end{center}
\end{table}

\section{Trigger-object-level object analysis in ATLAS and turbo stream in LHCb}

\begin{figure}[htbp]
\centering
\includegraphics[width=0.45\textwidth]{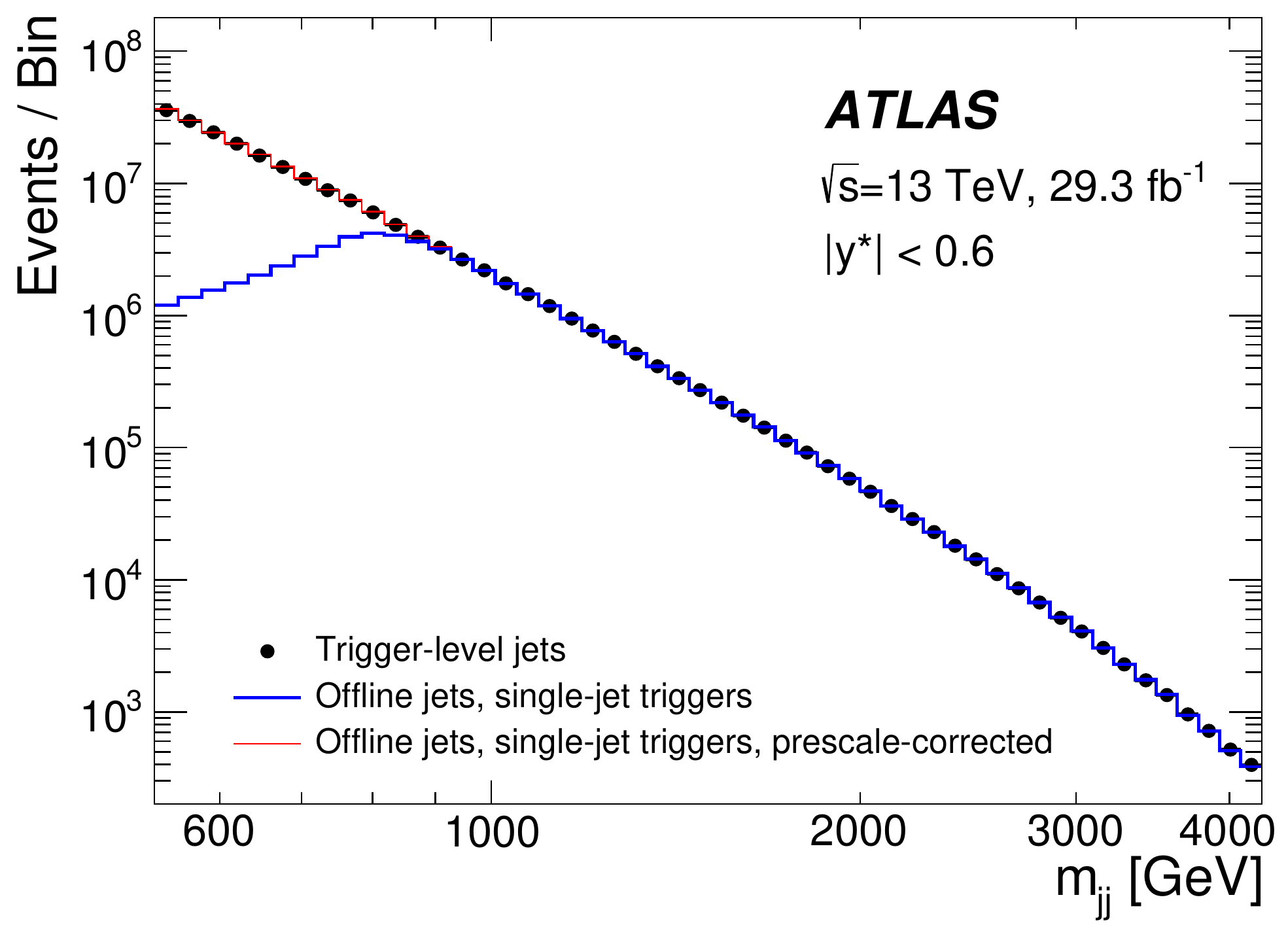}
\includegraphics[width=0.36\textwidth]{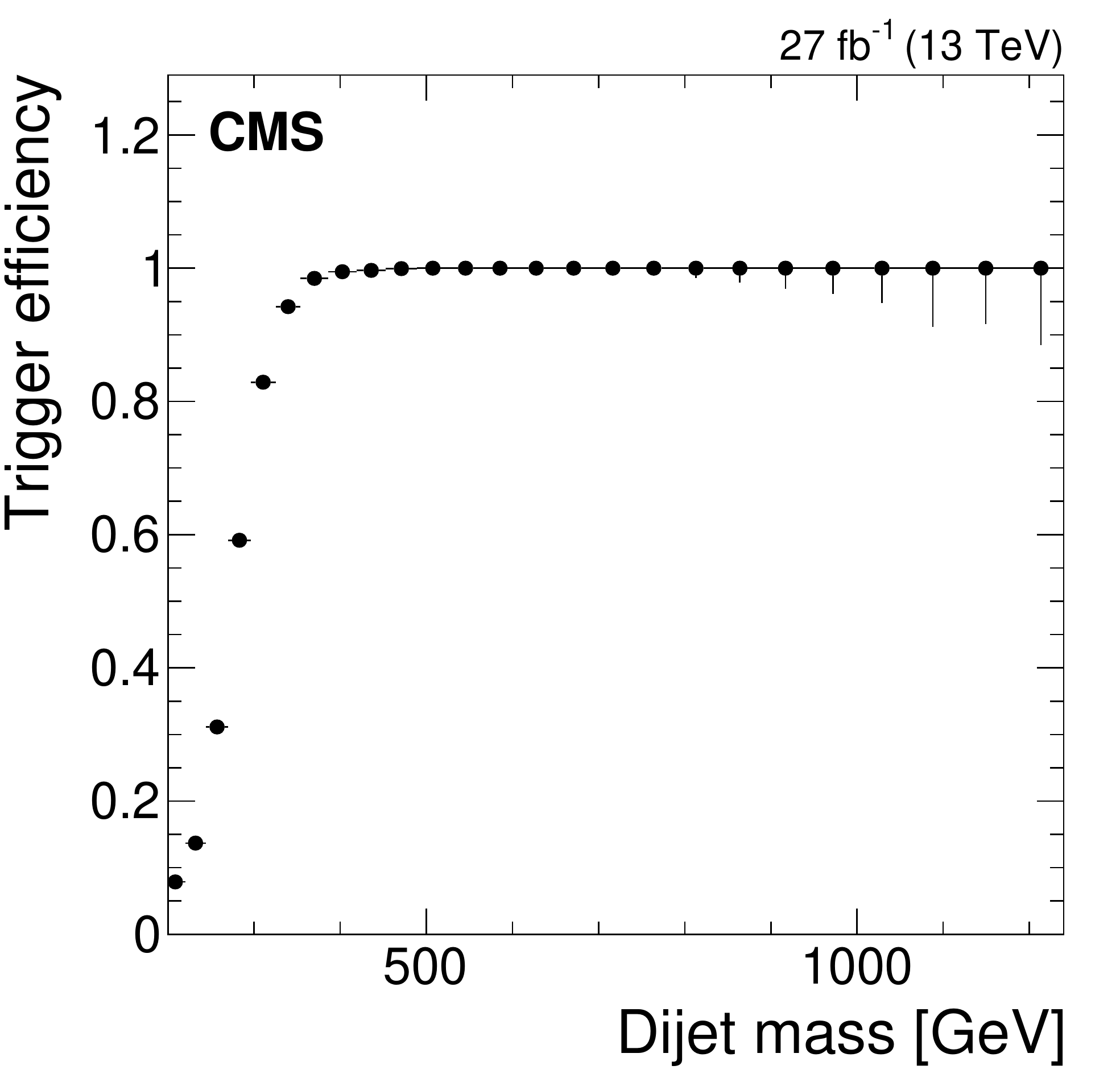}
\caption{Comparison between event acceptance for offline jets and
  trigger-level jets for ATLAS (left) and calo-scouting trigger efficiency for CMS (right).}
\label{fig:triggers}
\end{figure}

In ATLAS, the trigger-object-level analysis (TLA) approach allows jet
events to be recorded at a peak rate of up to twice the total rate of
events using the standard approach, while using less than 1\% of the
total trigger bandwidth~\cite{TRIG-2016-01}. The HLT reconstructs
anti-$k_t$ $R = 0.4$ jets from groups of contiguous calorimeter cells
(topological clusters), in which each cell's inclusion is based on the significance of its energy deposit over
calorimeter noise~\cite{PERF-2014-07}. Trigger-level jets with
$p_\mathrm{T}>20$~GeV are stored, including a set of calorimeter variables
characterizing the jet~\cite{ATLAS-CONF-2015-029}, such as information
about the jet quality and structure. The size of these events is less
than 0.5\% of the size of full events. All events containing at least
one L1 jet with $E_\mathrm{T}>100$~GeV are selected and recorded in
the 2016 dataset. The gain in recorded events from ATLAS TLA is shown in Fig.~\ref{fig:triggers}.

At the beginning of LHC Run 2, LHCb, with its upgraded computing
infrastructure and a more efficient use of the Event Filter Farm (EFF)
storage of 5.2 PB, began providing resources for an online
reconstruction with a similar quality to that of the offline
reconstruction. This is achieved through real-time automated calculation of the final calibrations of the sub-detectors. In the \emph{turbo stream}, a
compact event record is written directly from the trigger and is
prepared for physics analysis by the Tesla application. Using the turbo
stream, LHCb was able to record $J/\psi\to\mu^+\mu^-$ candidates with invariant masses
$m(\mu^+\mu^-)$ within 150~MeV of the known value 
and measure the forward $J/\psi$ production cross
section~\cite{Aaij:2015rla}. The LHCb turbo stream was also used to
measure prompt charm production cross sections~\cite{Aaij:2015bpa}.

\section{Dijet resonance searches in CMS and ATLAS}
\label{dijetCMS}

CMS data collected in the calo-scouting stream in 2012 and 2016 were used to
perform searches for dijet resonances with masses in the range
0.5--1.6~\cite{Khachatryan:2016ecr} and 0.6--1.6~TeV~\cite{Sirunyan:2018xlo}, respectively. The
2016 search was carried out with 27~fb$^{-1}$ of pp collision events\footnote{The L1 $H_\mathrm{T}$
  triggers suffered an inefficiency in 9 fb$^{-1}$ of data at the end
  of the run.} satisfying the $H_\mathrm{T}>250$~GeV trigger
requirement. Geometrically close anti-$k_t$ $R=0.4$ jets are combined into two ``wide jets''
with a jet radius of $R=1.1$ and the dijet invariant mass is required
to be $m_\mathrm{jj}>0.49$~TeV. Background from $t$-channel dijet events is
suppressed by requiring $|\delta\eta_\mathrm{jj}|<1.3$. 

The jet energy scale of the HLT calorimeter jets was calibrated to be the same as the
jet energy scale of the offline PF jets using the monitoring dataset
and a dijet balance ``tag-and-probe'' method~\cite{Khachatryan:2016kdb}. 

\begin{figure}[htbp]
\centering
\includegraphics[width=0.45\textwidth]{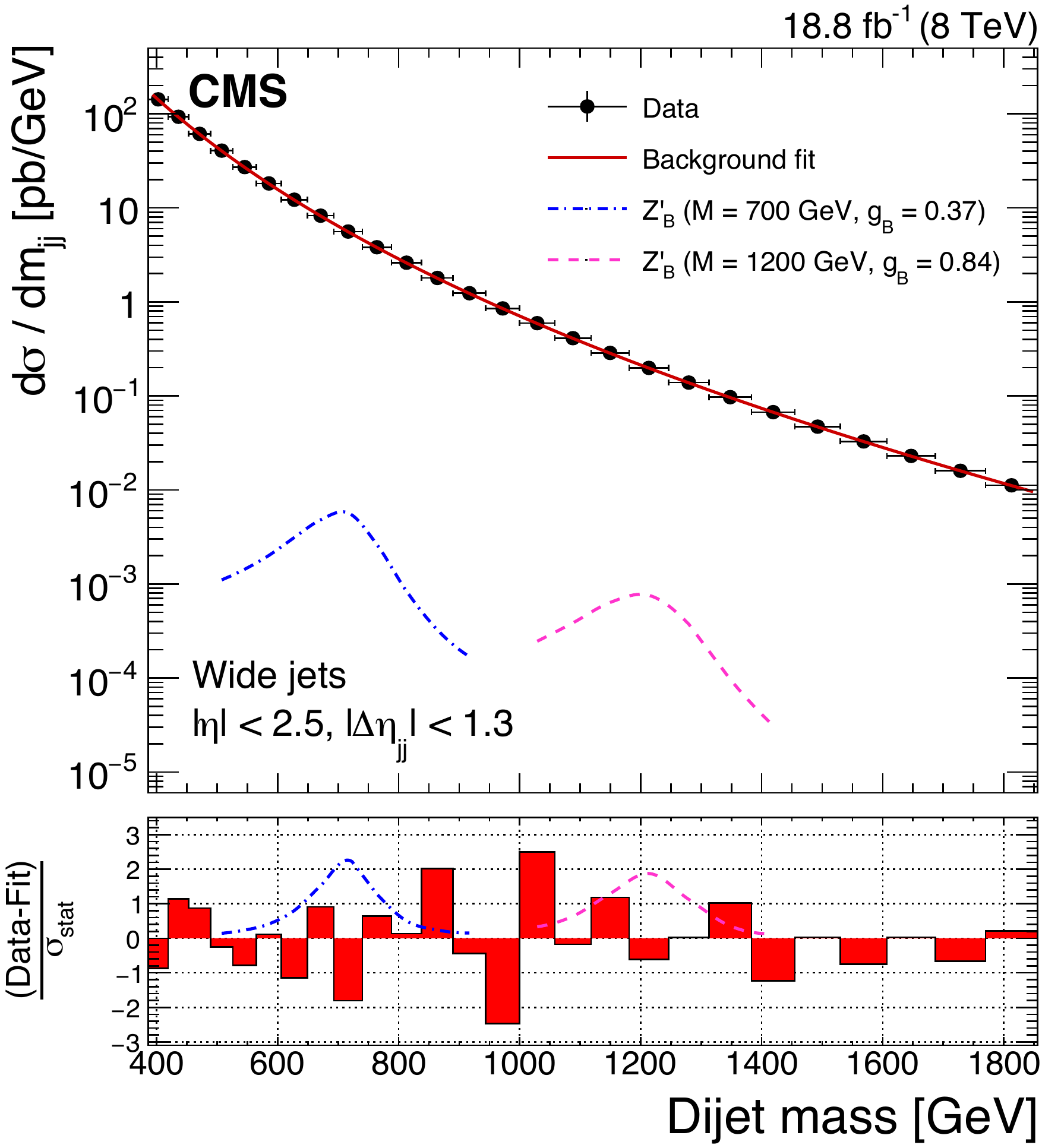}
\includegraphics[width=0.45\textwidth]{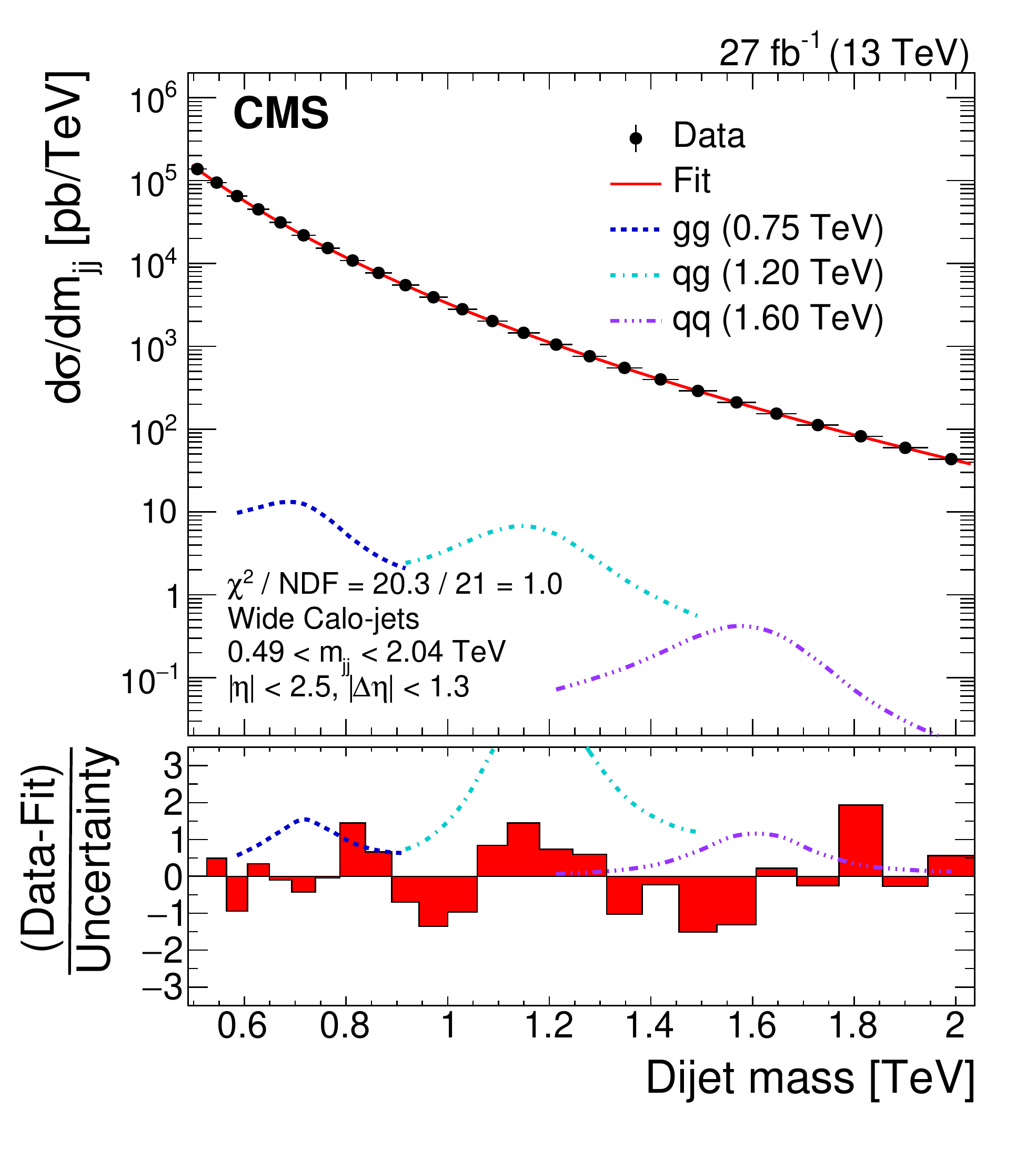}
\caption{Parametric fits of CMS low-mass calo-scouting 2012 data
  (left)~\cite{Khachatryan:2016ecr} and 2016 data (right)~\cite{Sirunyan:2018xlo}.}
\label{fig:fitsCMS}
\end{figure}

Fig.~\ref{fig:fitsCMS} shows the parametric fits to the scouting dijet mass spectra in both 2012 and 2016. 

The analogous dijet search in ATLAS used 29.3 fb$^{-1}$ of 2015 and
2016 data and required at least two trigger-level jets with leading
(subleading) $p_\mathrm{T} > 220$~($85$)~GeV and $|\eta|<2.8$. The analysis search for dijet
resonances with a masses between 450~GeV and 1800~GeV. To search for
resonances with $700 < m_\mathrm{jj} < 1800$~GeV, events are required to have $|y^{\ast}| < 0.6$, where $y^{\ast} = (y_1-y_2)/2$ and
$y_1$ and $y_2$ are the rapidities\footnote{The rapidity, $y$, is
  defined as $\frac{1}{2}\log[(E+p_z)/(E-p_z)]$} of the highest- and
second-highest-$p_\mathrm{T}$ trigger-level jets. To search for
lower-mass resonances, with $m_\mathrm{jj} > 450$~GeV, events with $|y^{\ast}| <
0.3$ are selected from the 2015 data sample.

The trigger-level jet energy and direction are corrected to those of
simulated particle-level jets built from stable particles, excluding
muons and neutrinos. The custom calibration recipe includes the standard calibrations applied to
offline jets as well as any residual difference between trigger-level jets and offline jets is accounted
for in a dedicated trigger-to-offline correction, based on the $p_\mathrm{T}$ response and derived
from data in bins of jet $\eta$ and $p_\mathrm{T}$. After the full calibration procedure, the energy of trigger-level jets is
equivalent to that of offline jets to better than 0.05\% for invariant
masses of 400~GeV.

\begin{figure}[htbp]
\centering
\includegraphics[width=0.65\textwidth]{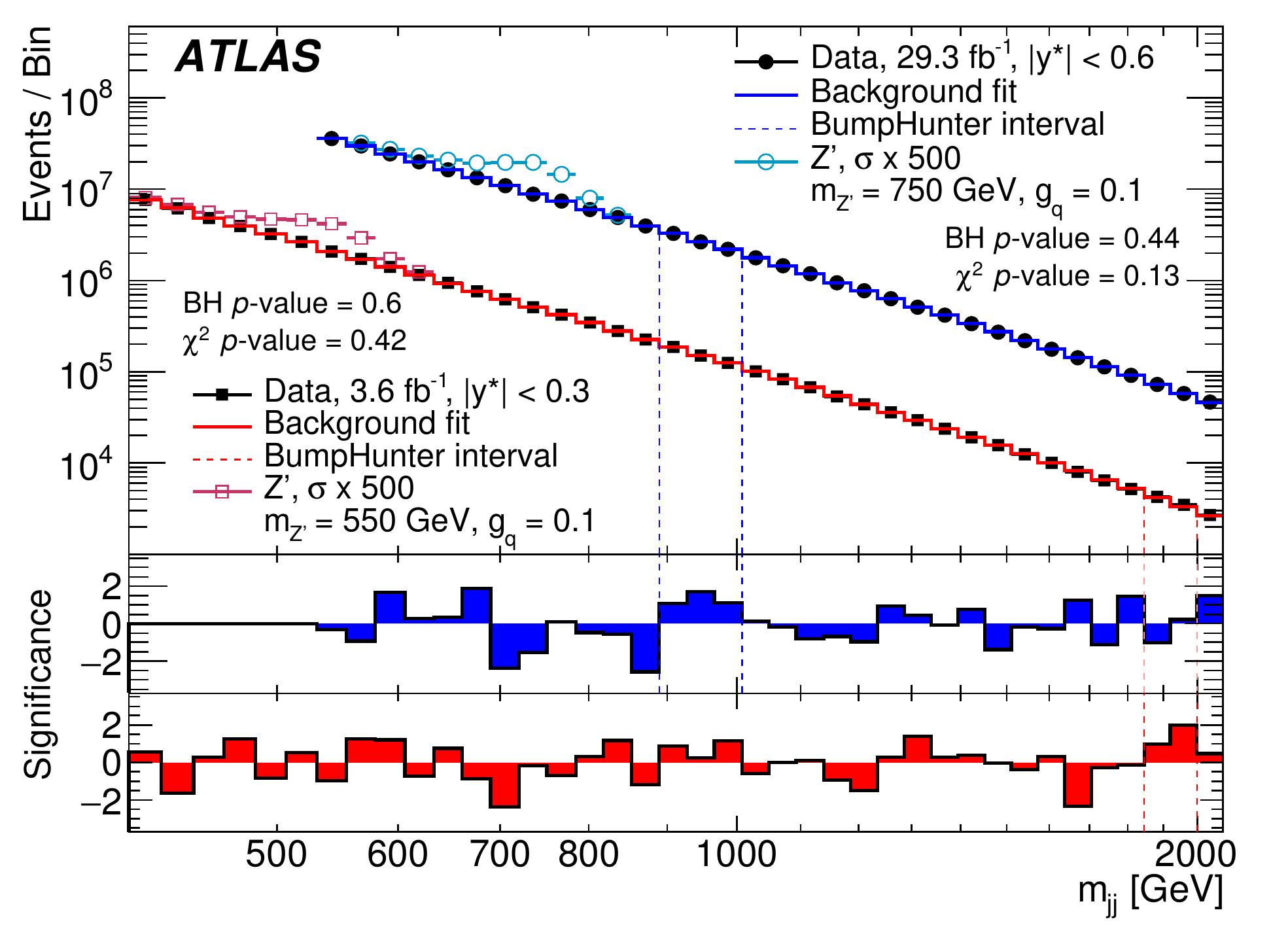}
\caption{Sliding window fits of ATLAS trigger-object-level analysis in 2015 and 2016 data~\cite{Aaboud:2018fzt}.}
\label{fig:fitsATLAS}
\end{figure}

Finally, the SM background distribution is determined
using a sliding-window fit~\cite{EXOT-2016-21} as shown in Fig.~\ref{fig:fitsATLAS}.

Upper limits on the coupling as a function of mass for a model of a
leptophobic $\mathrm{Z}^{\prime}$ resonance with a universal quark coupling,
$g^\prime_\mathrm{q}$~\cite{Dobrescu:2013coa,Abercrombie:2015wmb,Boveia:2016mrp}
are derived for both analyses. The interaction Lagrangian for this
model is

\begin{align}
\mathcal{L}_{\mathrm{Z}^{\prime}} &= \sum_\mathrm{q} g^\prime_\mathrm{q} \mathrm{Z}^{\prime}_{\mu}\bar{\mathrm{q}}\gamma^{\mu} \mathrm{q} ~.
\end{align}

\begin{figure}[htbp]
\centering
\includegraphics[width=0.9\textwidth]{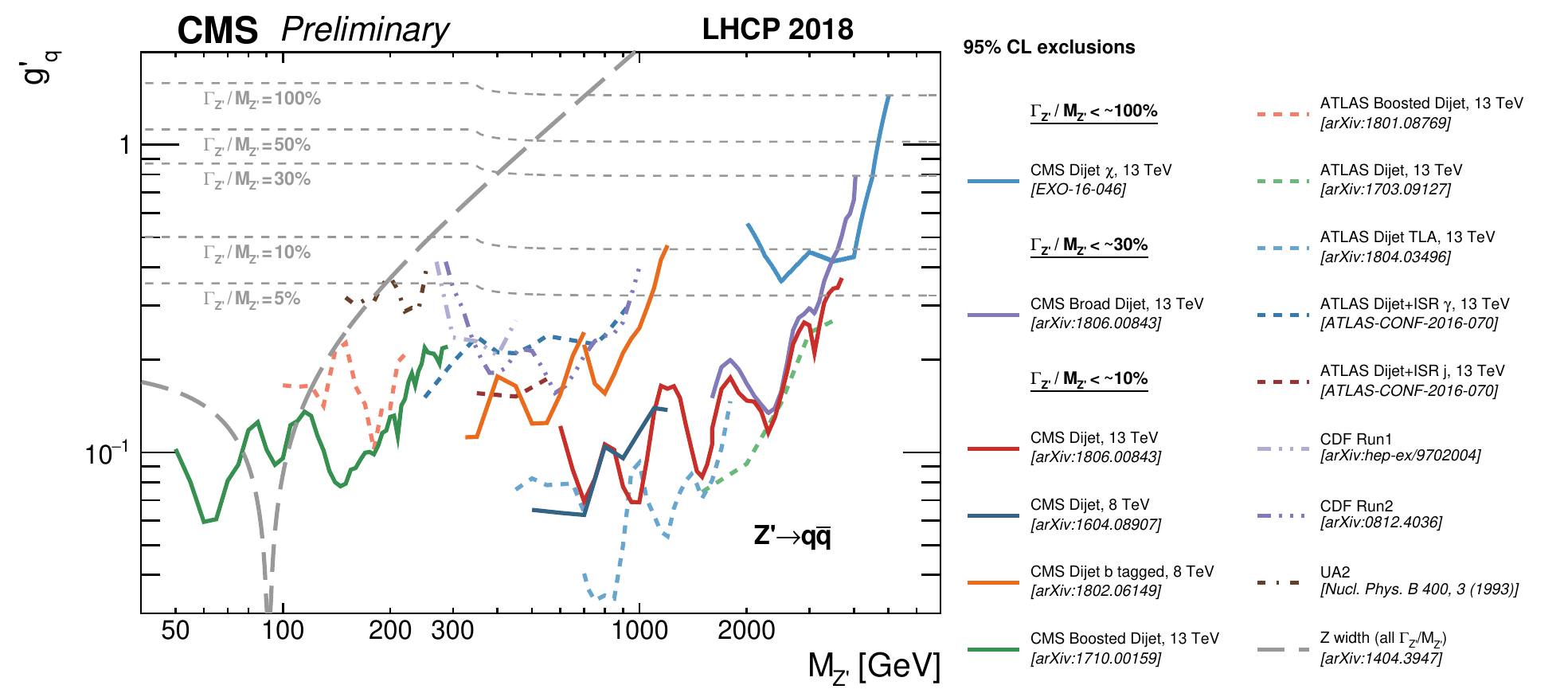}
\caption{Summary of 95\% CL upper bounds on the universal quark
  coupling for a leptophobic $\mathrm{Z}^{\prime}$ mediator.}
\label{fig:gqCMS}
\end{figure}

A summary of the constraints from these searches as well
as others is shown in Fig.~\ref{fig:gqCMS}. In particular, the impact
of the scouting and trigger-level analyses on the
$g^\prime_\mathrm{q}$ sensitivity can be seen for masses between $0.45$
and $1$~TeV. Over this mass range, the sensitivity to the coupling to
the universal quark coupling is improved by a factor of two or more
compared to pre-LHC searches.

\section{Summary}

Data scouting allows physics events to be collected at a rate
dramatically higher than what is nominally achievable with the
standard LHC trigger systems. It is implemented with no
changes needed to the basic software-based high-level-trigger
infrastructure and does not place an additional strain on the DAQ,
disk resources, or the reconstruction system. 

In CMS, the two scouting streams deployed for Run 2 of the LHC strike a
balance between specialization and versatility, with one stream
oriented towards dijet resonance searches and the other designed to
support arbitrary searches based on hadronic final states. 
The first physics results using the LHC Run 2 data scouting have been
published from CMS, ATLAS, and LHCb, and it is hoped that analyzers
will take full advantage of this tool to search in other previously
unexplored regions at the LHC.

\bibliography{template}

\end{document}